\documentclass[aps,floatfix, prl, reprint, superscriptaddress,nolongbibliography]{revtex4-2}
\usepackage[makeroom]{cancel}
\usepackage{bm}
\usepackage{amsmath,amssymb}

\usepackage{amsfonts}
\usepackage{siunitx}
\usepackage{graphicx}
\usepackage{physics}
\usepackage{mathtools}
\mathtoolsset{showonlyrefs=true,showmanualtags=true}

\newcommand{\be}{\begin{equation}}
\newcommand{\ee}{\end{equation}}

\usepackage{hyperref}
\usepackage{adjustbox} 
\usepackage[dvipsnames]{xcolor}

\usepackage{xcolor} % for defining colors
\usepackage{soul}   % for highlighting

\begin{document}

\title{Nonreciprocal inertial spin-wave dynamics in twisted magnetic nanostrips}

\author{Massimiliano d'Aquino}
\email{mdaquino@unina.it}
\affiliation{Department of Electrical Engineering and ICT, University of Naples Federico II, Naples, Italy}

\author{Riccardo Hertel}

\affiliation{Universit{\'e} de Strasbourg, CNRS, Institut de Physique et Chimie des Mat{\'e}riaux de Strasbourg, F-67000 Strasbourg, France}

\date{\today}
\begin{abstract}
We develop a theoretical framework for inertial spin-wave dynamics in three-dimensional twisted soft-magnetic nanostrips, where curvature and torsion couple with magnetic inertia to generate terahertz (THz) magnetic oscillations. The resulting spin-wave spectra exhibit pronounced nonreciprocity due to effective symmetry breaking arising from geometric chirality and inertial effects. We show that this behavior is governed by a curvature-induced geometric (Berry) phase, which we analytically capture through compact expressions for dispersion relations and spectral linewidths in both nutational (THz) and precessional (GHz) regimes. Topological variations, including Möbius and helical geometries, impose distinct wavenumber quantization rules, elucidating the role of topology in spin-wave transport.  These results position twisted magnetic strips as a viable platform for curvilinear THz magnonics and nonreciprocal spintronic applications.
\end{abstract}
\maketitle

%\section{Introduction}

The growing possibility to access the third dimension in the study of magnetic nanoscale systems has allowed to surpass the constraints imposed by traditional flat geometries\cite{fernandez-pacheco_three-dimensional_2017,gubbiotti_three-dimensional_2019,gubbiotti2025}. While thin-film magnetic systems have laid the foundation for much of modern spintronics\cite{Dieny2020opportunities} and magnonics\cite{kruglyak_magnonics_2010,chumak_roadmap_2022}, the transition to three-dimensional (3D) nanomagnetism introduces profound effects stemming from curvature, torsion, and topology\cite{hertel2013curvature,streubel2016magnetism,heyderman2021mesoscopic,sheka2022fundamentals,makarov_curvilinear_2022}.

In this context, the study of spin-wave propagation in geometrically complex nanostructures—such as twisted magnetic strips—has garnered increasing attention. These systems not only exhibit rich magnetization textures but also enable fundamentally new dynamic behaviors due to their inherent spatial asymmetry\cite{sheka2022fundamentals,makarov_curvilinear_2022}. Parallel to this, the detection of the emergence of terahertz (THz) nutation due to magnetic inertia\cite{Neeraj2021inertial} predicted more than a decade ago\cite{Ciornei2011magnetization,wegrowe2012magnetization} for ferromagnets has opened new routes in ultrafast magnetism and driven considerable research on inertial spin-wave dynamics\cite{kikuchi2015spin,giordano2020derivation,makhfudz2020nutation,lomonosov2021anatomy,cherkasskii2021dispersion,mondal2022inertial,titov2022nutation,daquino2023micromagnetic,cherkasskii2024inertial,rodriguez2024spin} following the pioneering experiments on planar magnetic nanostructures\cite{Neeraj2021inertial,unikandanunni2022inertial}. 

In this Letter, we demonstrate that excitation of inertial spin waves in curved geometries gives rise to novel dynamical phenomena occurring in the THz regime.
The work focuses on the inertial spin-wave dynamics in 3D twisted soft-magnetic nanostrips, where curvature and torsion couple with magnetic inertia to produce THz magnetic oscillations. We demonstrate that these inertial effects, amplified by the underlying 3D geometry, lead to a pronounced nonreciprocal response in the spin-wave spectrum. The combination of geometric chirality and inertial dynamics introduces an effective symmetry breaking, positioning these twisted architectures as promising candidates 
that pave the way to the field of curvilinear THz magnonics. 

The main result of this paper is a theoretical treatment with no adjustable parameters that allows quantitative understanding of the onset of geometric (Berry) phase\cite{pancharatnam1956generalized,wilczek1989geometric} in inertial spin wave dynamics, which is responsible for the aforementioned symmetry-breaking. Simple and compact formulas for dispersion relations and full width at half maximum (FWHM) spectral linewidths are derived both in the precessional (GHz) and in the nutational (THz) regimes and offer a complete picture of the influence of each single parameter on spin-wave dynamics. Application to nanostrips with different nontrivial topology such as M\"obius and helical strips reveals different wavenumber quantization rules, enlightening the role of topology in spin-wave dynamics similarly to what has been done for microwave resonators\cite{hamilton2021absorption} and recently led to the observation of optical Berry phase in Möbius-strip microcavities\cite{wang2023experimental}.

%\section{Inertial spin wave dynamics}

We consider a magnetic ultrathin nanostrip with a thickness of a few nanometers occupying a region $\Omega$ of volume $V$. Magnetization dynamics is governed by the inertial Landau-Lifshitz-Gilbert (iLLG) equation\cite{Ciornei2011magnetization,daquino2023micromagnetic}: 
\begin{equation}
\frac{\partial \bm m}{\partial t}=-\bm m\times \left(\bm  h_\mathrm{eff}-\alpha \frac{\partial \bm m}{\partial t} -\xi \frac{\partial^2 \bm m}{\partial t^2}\right)\quad, \label{eq:iLLG} 
\end{equation}
where magnetization $\bm m(\bm r,t)$ is normalized by the saturation magnetization $M_s$, time is measured in units of $(\gamma M_s)^{-1}$ ($\gamma$ is the absolute value of the gyromagnetic ratio), $\alpha$ is the Gilbert damping constant, $\xi=(\gamma M_s t_\mathrm{in})$ measures the strength of inertial effects ($t_\mathrm{in}$ is the physical timescale of inertial effects, in the order of fractions of picosecond, which implies $\xi\sim10^{-2}$), and $\bm h_\mathrm{eff}$ is the effective field $\bm h_\mathrm{eff} = \bm \ell_\mathrm{ex}^2\nabla^2\bm m+\bm h_\mathrm{m}+K_\mathrm{an}(\bm m \cdot\bm e_\mathrm{an})\bm e_\mathrm{an}+\bm h_\mathrm{a}$
which arises from the negative variational derivative of the Gibbs-Landau micromagnetic free energy functional $g[\bm m,\bm h_\mathrm{a}]$ and takes into account exchange ($\ell_\mathrm{ex}=\sqrt{(2A_\mathrm{ex})/(\mu_0 M_s^2)}$ and $A_\mathrm{ex}$ are the exchange stiffness and length, respectively), magnetostatic, magneto-crystalline (uniaxial) anisotropy ($K_\mathrm{an}$ is the anisotropy constant) and Zeeman interactions. Magnetization is assumed to satisfy the natural conditions $\partial\bm m/\partial\bm n=0$ on the boundary $\partial\Omega$ which imply the absence of surface anisotropy. The magnetostatic field $\bm h_\mathrm{m}$ is the solution of magnetostatic Maxwell's equations that can be expressed via the Green function as linear integral operator $\mathcal{N}$ acting on the magnetization unit-vector $\bm h_m=-\mathcal{N}\bm m$.
Magnetization dynamics expressed by eq.\eqref{eq:iLLG} is constrained on the unit-sphere $|\bm m|^2=1$ at every location $\bm r\in\Omega$ due to the fundamental micromagnetic constraint\cite{daquino2024midpoint}. 

\begin{figure}[t]
    \centering
    \includegraphics[width=0.9\linewidth]{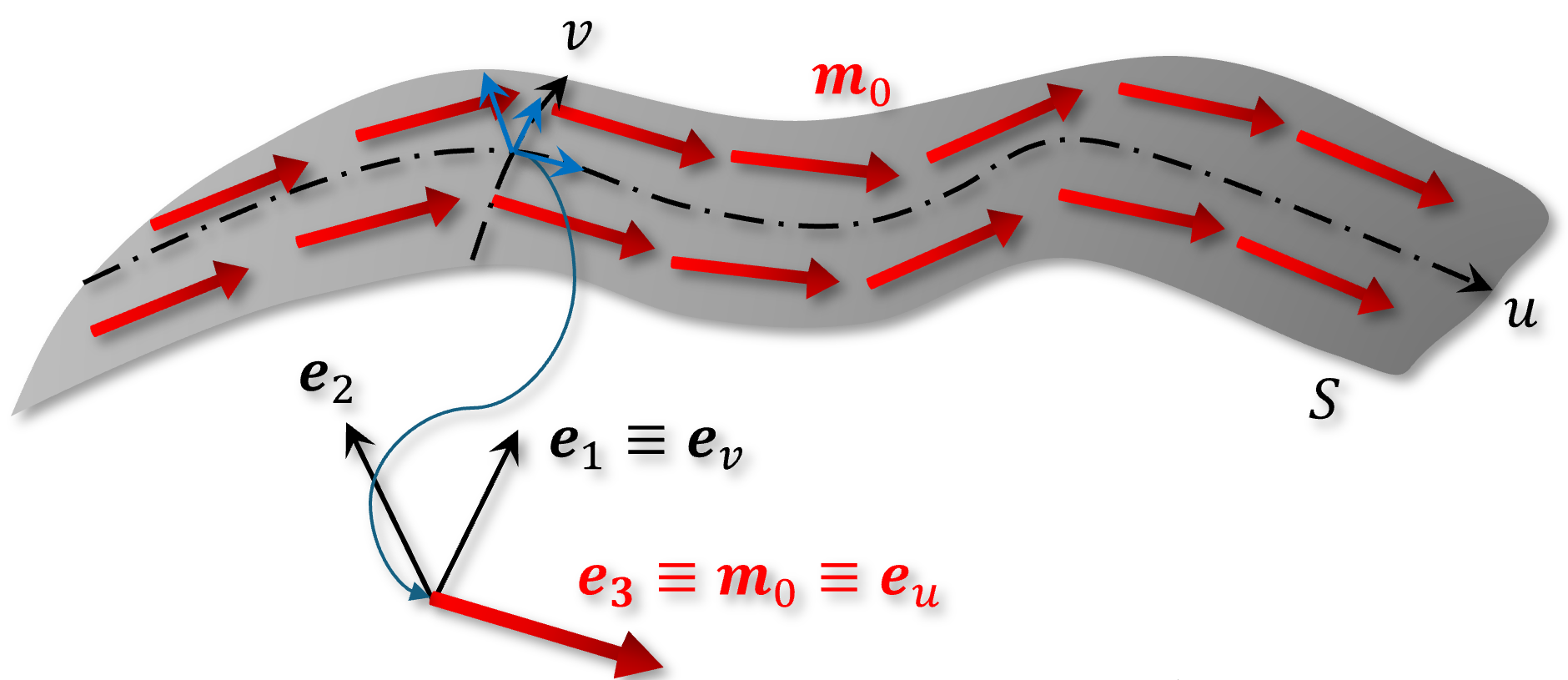}
    \caption{Sketch of a curved ultrathin strip. $(\bm e_1, \bm e_2, \bm e_3)$ is the local reference frame induced by the parametric representation $\bm r_S(u,v)$ as function of the curvilinear coordinates $(u,v)$.}
    \label{fig:curved strip}
\end{figure}

Small magnetization oscillations $\delta \bm m(\bm r,t)$ around an equilibrium configuration $\bm m_0(\bm r)$ are described by the linearized iLLG equation:
\begin{equation}
\frac{\partial \delta\bm m}{\partial t}=-\bm m_0\times\left( \delta\bm h_\mathrm{eff} - h_0\delta\bm m 
-\alpha\frac{\partial\delta\bm m}{\partial t}-\xi\frac{\partial^2\delta\bm m}{\partial t^2} \right)\,, \label{eq:linearized iLLG} 
\end{equation}
where $\delta \bm h_\mathrm{eff}[\delta\bm m]=\ell_\mathrm{ex}^2\nabla^2 \delta\bm m+ h_\mathrm{m}[\delta\bm m] +K_\mathrm{an}( \delta\bm m \cdot\bm e_\mathrm{an})\bm e_\mathrm{an}$ groups terms in the effective field that are linear with respect to magnetization and $h_0(\bm r)=\bm h_\mathrm{eff}[\bm m_0]\cdot \bm m_0$ is the component of the effective field along the magnetization at the equilibrium. Due to the unit-sphere constraint, the small oscillation dynamics occurs in the plane point-wise transverse to the equilibrium magnetization such that $\delta\bm m\cdot\bm m_0=0$.
For this reason, the vector field $\delta\bm m$ can be represented using an orthogonal (positively-oriented) local basis $(\bm e_1,\bm e_2, \bm e_3)$
where $\bm e_3=\bm m_0(\bm r)$ and the remaining vectors $\bm e_1,\bm e_2$ can be defined up to a rotation around $\bm e_3$ at any location $\bm r\in\Omega$. Thus, remembering the aforementioned fundamental constraint, one has $\delta\bm m(\bm r,t)=\delta m_1(\bm r,t) \bm e_1(\bm r) + \delta m_2(\bm r,t) \bm e_2(\bm r)$.
In this respect, one can express magnetization oscillation $\delta\bm m$ as superposition of eigenmodes\cite{daquino_novel_2009}, as $\delta \bm m(\bm r,t)=\Re{\sum_h a_h \left(\varphi_{h1}(\bm r)\bm e_1(\bm r)+ \varphi_{h2}(\bm r)\bm e_2(\bm r)\right)e^{i\omega_h t}}$,
where $a_h$ is a constant (complex) coefficient and each eigenmode $\bm \varphi_h$ has been decomposed along the local unit-vectors $\bm e_1,\bm e_2$.
Now, in order to derive a simpler dynamical equation for the magnetization $\delta\bm m$, we assume that the nanostrip has a thickness comparable to the exchange length and therefore magnetization does not change appreciably along the thickness direction. This means that $\delta \bm m(\bm r,t)$ can be defined on the surface $S$ that is expressed through its parametric representation $\bm r=\bm r_S(u,v)$ where $u,v$ are the curvilinear coordinates along the axis (with length $L$) and the width $w$ of the strip, respectively and $\bm e_u(u,v), \bm e_v(u,v)$ are the associated unit-vectors, as depicted in Fig.\ref{fig:curved strip}. 

\begin{figure}[t]
    \centering
    \includegraphics[width=0.6\linewidth]{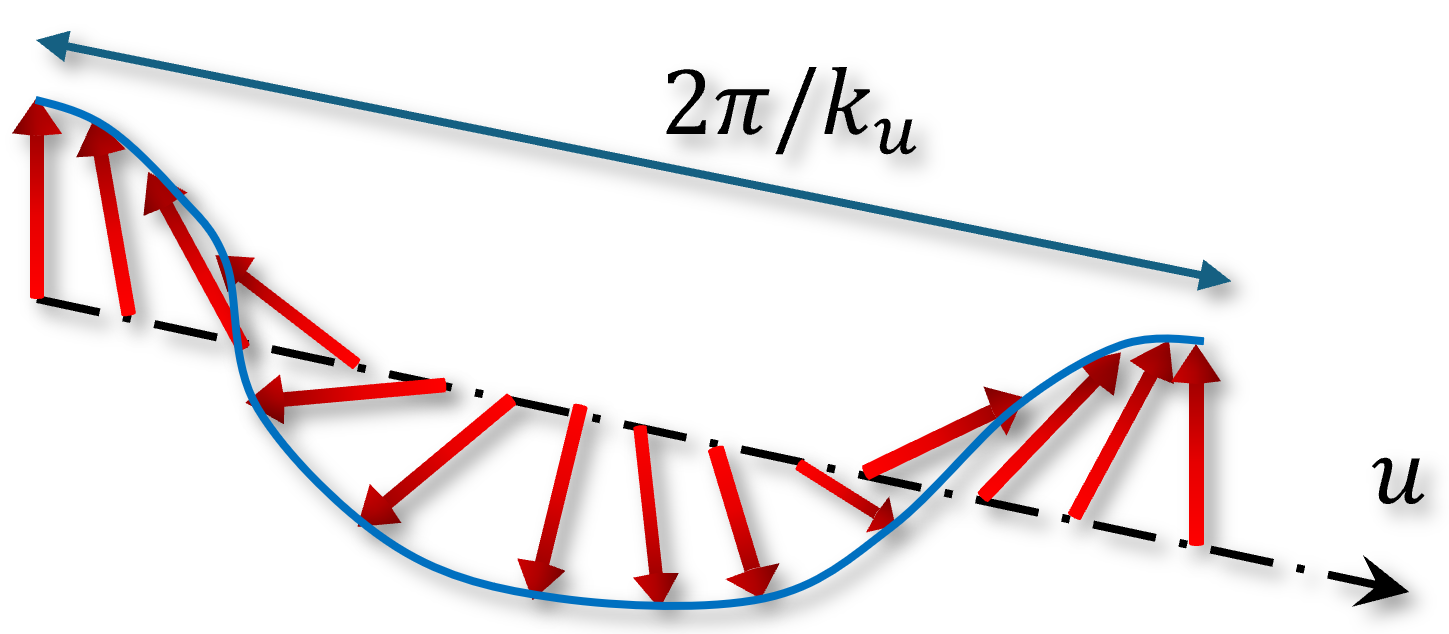}
    \caption{Helical spin wave rectified along the curvilinear abscissa $u$ of the strip axis.}
    \label{fig:helical spin wave}
\end{figure}

%\section{Geometric phase in helical spin waves}

We assume that the strip is much longer than wide, namely $L\gg w$, and that the anisotropy unit-vector is directed along the axis of the strip, $\bm e_\mathrm{an}=\bm e_u$. In this situation, due to shape (magnetostatics) and crystalline anisotropy, the equilibrium magnetization is expected to be quasi-tangential for not too large curvature\cite{sheka2022fundamentals} to the surface $S$ and mostly oriented along the axis of the surface $\bm m_0(\bm r=\bm r_S(u,v)) \approx \bm e_3(u,v) = \bm e_u(u,v)$.
This means that one can also pose $\bm e_1(u,v)=\bm e_v(u,v) \quad,\quad \bm e_2(u,v)=\bm e_3\times\bm e_1$.
The assumption $w\ll L$ also implies that the dependence on the transverse coordinate $v$ can be neglected in the local basis unit-vectors $\bm e_q\approx\bm e_q(u) \,,\, q=1,2,3$.
Now we consider a single normal mode $\bm \varphi_h(\bm r)e^{i\omega_h t} =  \bm \varphi_h(\bm r=\bm r_S(u,v))e^{i\omega_h t} = \bm \varphi_h(u,v)e^{i\omega_h t}$
expressed with little abuse of notation as function of the surface coordinates $(u,v)$. By Fourier-transforming with respect to the coordinate $u$
 and averaging along the width $w$, one obtains:
\begin{align}
   &\bm\varphi_h(u,v)e^{i\omega_h t}\approx \!\!\sum_{q=1,2}\!\! \frac{1}{2\pi} \int_{-\infty}^{+\infty} \!\!\!\!\hat{\Phi}_{hq}(k_u) e^{i(\omega_h t-k_u u)} \, dk_u  \bm e_q(u), 
\end{align}
where the components $\hat{\Phi}_{hq}(k_u)=\frac{1}{w}\int_{-w/2}^{w/2} \hat{\varphi}_{hq}(k_u,v)e^{i\omega_h t} \,dv$ (notations $\hat{\Phi}_{hq},\hat{\varphi}_{hq}$ denote Fourier transforms) only depend on the wavenumber $k_u$ and determine a vector field $\hat{\bm \Phi}_h=\hat{\Phi}_{h1}(k_u)\bm e_1(u) + \hat{\Phi}_{h2}(k_u)\bm e_2(u)$ that lies in the plane $\bm e_1,\bm e_2$ perpendicular to the wave vector $k_u\bm e_3$.

This offers an interesting physical interpretation, in that the vector $\hat{\bm \Phi}_h e^{i(\omega_h t-k_u u)}$ describes a helical spin wave (circularly polarized in the plane $(\bm e_1,\bm e_2)$ twisting along the length of the strip with chirality defined by the sign of its wavenumber $k_u$ (see the sketch in Fig.\ref{fig:helical spin wave}).

In order to quantitatively describe the effect of the curvature on spin waves propagation along the strip, we assume that the parametric representation $\bm r_S(u,v)$ is that of a ruled surface\cite{docarmo2016differential}, namely $\bm r_S(u,v)=\bm r_a(u)+v\, \bm r_r(u)$,
where the vector $\bm r_a(u)$ defines the axis curve of the surface $S$ and the vector $\bm r_r$ is the so-called ruling of the surface $S$. 
One can easily compute the triple $(\bm e_1,\bm e_2, \bm e_3)$ as $\bm e_1=\frac{d\bm r_S/dv}{||d\bm r_S/dv||} \,,\, \bm e_2=\bm e_3\times \bm e_1 \,,\, \bm e_3=\frac{d\bm r_S/du}{||d\bm r_S/du||}$.

Let the axis $\bm r_a(u)$ have tangent, normal and binormal unit-vectors given by $(\bm t,\bm n, \bm b)$, respectively, which define a positively-oriented triple. The ruled surface $S$ can be then characterized by two scalar functions $\kappa(u)$ and $\tau(u)$ being the curvature and the torsion of the axis line.

In this respect, the ruling $\bm r_r(u)$ can be expressed as $\bm r_r(u)=\cos\left(\theta(u)\right)\bm n(u) +\sin\left(\theta(u) \right) \bm b(u)$,
that describes a continuous rotation (twisting) of $\bm n,\bm b$ around the tangent unit-vector $\bm t$ with elementary rotation $d\theta = \tau du$, which implies the twist angle
\begin{equation}
    \theta(u)=\int_{0}^{u} \tau(u') du' + C \quad, \label{eq:twist angle}
\end{equation}
being $C$ an arbitrary constant phase-shift. 

Thus, the effect of the torsion of the strip can be understood as a geometrical (Pancharatnam-Berry) phase\cite{pancharatnam1956generalized,wilczek1989geometric} accumulation $\theta(u)$ along the strip that arises from a continuous rotation of the spin-wave polarization plane $(\bm e_1(u),\bm e_2(u))$. In this respect, one could imagine that the spin wave mode at the same frequency $\omega_h$ in a flat strip would appear as $\bm \varphi_\mathrm{h,flat}(u,v)\approx\hat{\bm \Phi}_h' e^{i(\omega_h t-k_u u+\theta(u))}$,
where the change in wave polarization is denoted with $\hat{\bm \Phi}_h'$ and the twist angle is subtracted to the linearly-varying phase of the spin wave. 

%\section{1D spin wave dynamics}

We now use the setting introduced in the previous sections to derive a simple equation for determining helical spin waves dynamics in a generic ruled surface $S$ defined as $\bm r_S(u,v)=\bm r_a(u)+v\bm r_r(u)$. 
To this end, we consider the linearized iLLG equation \eqref{eq:linearized iLLG} rewritten using the one-dimensional assumption 
for the local reference frame $(\bm e_1,\bm e_2,\bm e_3)$. 
For the sake of simplicity, we consider magnetostatics within the approximation of the ultrathin strip $d\sim\ell_\mathrm{ex}$. Moreover, since $L\gg w$, the demagnetizing field along the axis of the strip is negligible. These assumptions imply that the magnetostatic operator can be approximated by a local\cite{gioia1997micromagnetics,slastikov2005micromagnetics,kohn2005another,gaididei2017magnetization} second-order tensor $N$ that has nonzero components (i.e. the demagnetizing factors) only along the transverse directions, namely $\mathcal{N}\varphi(u) \approx N\cdot\bm\varphi=N_1\varphi_1\bm e_1+N_2\varphi_2\bm e_2$.
For nanostrips with rectangular cross-section of width $w$ and height $d$, the demagnetizing factors $N_1,N_2$ are those reported in ref\cite{aharoni1998demagnetizing}.
Then, we make a further simplifying assumption considering that the internal field at the equilibrium $h_0$ is uniform along the strip (this will also include the anisotropy along the axis through the coefficient $K_\mathrm{an}$, if present). The latter assumption implies a spectral shift that will concur to determine the fundamental mode resonance frequency, resulting in the following equation:
\begin{equation}
     -i\omega\bm e_3(u)\times \bm\varphi(u)
     =[-\ell_\mathrm{ex}^2\nabla^2 + N + (h_0-\xi\omega^2+i\omega\alpha)\mathcal{I}] \bm \varphi(u) \,.
\end{equation}
To keep things as simple as possible, we assume that the ruled surface has constant curvature $\kappa$ and torsion $\tau$. 
Then, expressing the second derivative along the axis of the strip in the local $(\bm e_1,\bm e_2, \bm e_3)$ reference, the latter equation can be decomposed on the transverse plane as:
\begin{align}
    i\omega \varphi_2&=-\ell_\mathrm{ex}^2 \dv[2]{\varphi_1}{u} + \ell_\mathrm{ex}^2(\kappa^2+\tau^2) \varphi_1+  \\ &+2\tau\ell_\mathrm{ex}^2\dv{\varphi_2}{u} + (\omega_{01}-\xi\omega^2+i\omega\alpha) \varphi_1 \,, \label{eq:QEP1}\\
    -i\omega \varphi_1&=-\ell_\mathrm{ex}^2 \dv[2]{\varphi_2}{u} + \ell_\mathrm{ex}^2\tau^2 \varphi_2 +  \\
    &-2\tau\ell_\mathrm{ex}^2\dv{\varphi_1}{u} + (\omega_{02}-\xi\omega^2+i\omega\alpha) \varphi_2 \,,   \label{eq:QEP2} 
\end{align}
where $\omega_{01}=h_0+N_1$ and $\omega_{02}=h_0+N_2=h_0+1-N_1$.
By making the Fourier ansatz (plane-wave-like) $\varphi_q(u)=\hat{\varphi}_q e^{-i k u}, q=1,2$, 
and denoting with $\omega_{11}(k)=\omega_{01} + \ell_{\mathrm{ex}}^2(k^2 + \tau^2 + \kappa^2)$ and $\omega_{22}(k)=\omega_{02} + \ell_{\mathrm{ex}}^2(k^2 + \tau^2)$, 
one has that nontrivial solutions of the problem require that
\begin{align}
   &P(\omega)= \xi ^2\,\omega^4-2i\alpha\xi \omega^3-\left(1+\alpha^2+\xi(\omega_{11}+\omega_{22})\right)\,\omega^2 +\\
    +&[i\alpha(\omega_{11}\!+\!\omega_{22})\!-\!4k\tau \ell_\mathrm{ex}^2] \omega\!+\!\omega_{11}\,\omega_{22}-(2k\tau\ell_\mathrm{ex}^2)^2\!=\!0 \,. \label{eq:characteristic polynomial 4degree}   
\end{align}
The four roots $\omega=\pm\omega_N+i\Delta\omega_N,\pm\omega_P+i\Delta\omega_P$ of the characteristic polynomial  \eqref{eq:characteristic polynomial 4degree} are the (complex) eigenfrequencies associated with nutation and precession  inertial spin-wave eigenmodes, respectively. In this respect, the real parts $\omega_N,\omega_P$ correspond to the natural oscillation frequencies, whereas the imaginary parts $\Delta\omega_N,\Delta\omega_P$ are associated with temporal decays (full width at half maximum (FWHM) linewidths are $2\Delta\omega_N,2\Delta\omega_P$). By using appropriate perturbation theory, one obtains the following expressions that are accurate in the range $|k\ell_\mathrm{ex}/\pi|\leq 1$ that extends to large wavenumbers corresponding to ultra-short spin waves:

%\begin{widetext}    
\onecolumngrid
\begin{align}
  \omega_N&
  \approx\sqrt{\frac{1+\xi(\omega_{11}+\omega_{22}) +\sqrt{1+2\xi(\omega_{11}+\omega_{22}) }}{2\,\xi ^2}}+ (2k\tau\ell_\mathrm{ex}^2)\,\left(1-\xi \,\left(\omega_{11}+\omega_{22}\right)\right) \label{eq:dispersion relation nutation} \\
    \omega_P&\approx 
\frac{\sqrt{\omega_{11}\,\omega_{22}}}{\sqrt{1+\xi(\omega_{11}+\omega_{22}) }}-(2k\tau\ell_\mathrm{ex}^2)\,\left(1-\xi \,\left(\omega_{11}+\omega_{22}\right)\right) \label{eq:dispersion relation precession} \\
    \Delta\omega_N&\approx\left(1+\frac{1}{\sqrt{1+2\xi(\omega_{11}+\omega_{22}) }} \right)\,\frac{\alpha}{2\xi} +2k\tau\ell_\mathrm{ex}^2\left(1-\xi(\omega_{11}+\omega_{22}) \right)\,,  \label{eq:nutation linewidth} \\
    \Delta\omega_P&\approx\left(1-\frac{1}{\sqrt{1+2\xi(\omega_{11}+\omega_{22}) }}\right)\,\frac{\alpha}{2\xi} +2k\tau\ell_\mathrm{ex}^2\left(1-\xi(\omega_{11}+\omega_{22}) \right)\,, \label{eq:precession linewidth}\\    
\end{align}

%\end{widetext}
\twocolumngrid
Equations \eqref{eq:dispersion relation nutation}-\eqref{eq:dispersion relation precession} define the nutation  and precession dispersion relation branches of the spin wave eigenmodes on the twisted strip $S$ (we remark that $\omega_{11}(k),\omega_{22}(k)$ are functions of the wavenumber $k$). The formulas \eqref{eq:dispersion relation nutation}-\eqref{eq:dispersion relation precession} are simple and compact and allow one to understand the influence of each parameter on the spin wave dynamics. 

First, one can see that the resonance frequency of the fundamental nutation mode is given by:
\begin{align}
    \omega_N\overset{k\rightarrow 0}{=}
  \frac{1}{\xi}+\frac{1+2h_0 + \ell_{\mathrm{ex}}^2(2\tau^2 + \kappa^2)}{2}\,, \label{eq:nutation fundamental frequency}
\end{align}
and occurs in the THz range ($\xi \sim 4.5\times 10^{-2}$ in permalloy means $\gamma M_s/(2\pi\xi)\sim 630$ GHz).
Analogously, one has for the precession (Kittel)  fundamental mode:
\begin{equation}
\omega_P\!\overset{k\rightarrow 0}{=}\!\sqrt{\frac{[h_0\!+\!N_1\!+\!\ell_\mathrm{ex}^2(\tau^2+\kappa^2)](h_0\!+\!N_2\!+\!\ell_\mathrm{ex}^2\tau^2)}{1+\xi[1+2h_0 + \ell_{\mathrm{ex}}^2(2\tau^2 + \kappa^2)]}} . \label{eq:FMR frequency}
\end{equation} 
Equations \eqref{eq:nutation fundamental frequency}-\eqref{eq:FMR frequency} show that both fundamental precession and nutation resonances are affected by the curvature $\kappa$ and torsion $\tau$ of the strip which produce frequency blueshifts. Moreover, concerning the precession dispersion relation, it is evident from formula \eqref{eq:FMR frequency} that the influence of inertia can only induce frequency redshifts compared to the classical non-inertial case $\xi=0$.
Then, it is apparent that both nutational and precessional branches of the dispersion relation are not symmetric in the presence of a twisted surface with $\tau\neq 0$, thus indicating nonreciprocity of spin waves propagation. Conversely, the curvature $\kappa$ just produces a frequency shift without breaking the symmetry (reciprocity) of the dispersion relation. By observing the second terms proportional to $\tau$ in eqs.\eqref{eq:dispersion relation nutation}-\eqref{eq:dispersion relation precession}, we notice that the twisting produces the same asymmetry but with opposite signs in precessional and nutational spin waves, respectively. This can be seen as an additional fingerprint to discriminate inertial dynamics in possible experiments.

% We also note that, when $\kappa=\tau=0$, one strikingly recovers the classical dispersion relation of exchange spin waves for purely flat strips.

The group velocity can be computed as $v_g(k)=\gamma M_s \partial\omega/\partial k$ using eqs.\eqref{eq:dispersion relation nutation}-\eqref{eq:dispersion relation precession} in their range of validity $|k\ell_\mathrm{ex}/\pi|\leq 1$, and it is apparent that both precessional and nutational spin waves clearly exhibit asymmetry (again, with opposite signs) due to the term $\mp2\tau\ell_\mathrm{ex}^2$, respectively. 

Moreover, in the large wavenumber limit $|k\gg \pi/\ell_\mathrm{ex}|$, it can be easily seen that both branches of the dispersion relation admit the same ultimate speed limit for inertial spin waves\cite{daquino2023micromagnetic}. This can be understood by observing that, for large $k$, one can put $\omega_{11}+\omega_{22}\sim 2\ell_\mathrm{ex} k^2$, $\omega_{11}\omega_{22}\sim (\ell_\mathrm{ex}^2 k^2)^2$ in eq.\eqref{eq:characteristic polynomial 4degree} with $\alpha=0$, assume $\omega\sim k^2$ and neglect terms of order smaller than $k^4$ in the solution of \eqref{eq:characteristic polynomial 4degree}, which yields $v_g(k\rightarrow\pm\infty)=\pm\gamma M_s\frac{\ell_\mathrm{ex}}{\sqrt{\xi}}=\pm\frac{\ell_\mathrm{ex}}{t_\mathrm{in}}$.

We now look at the temporal decay constants \eqref{eq:nutation linewidth}-\eqref{eq:precession linewidth}, stressing that the same asymmetry appears due to the nonzero torsion as in the dispersion relations. In addition,   
we note that $\Delta\omega_N$ has a maximum close to $k=0$ whereas $\Delta\omega_P$ has a minimum. This can be easily understood by using the Taylor expansion $1\pm 1/\sqrt{1+x}\approx 1\pm 1\mp x/2\,,\, x\ll 1$ in formulas \eqref{eq:nutation linewidth}-\eqref{eq:precession linewidth} and remembering that $\omega_{11}+\omega_{22}$ increases as a function of $k^2$. In fact, this leads to $\Delta\omega_N(k\ll 1)\approx \frac{\alpha}{\xi}(1-\xi\frac{\omega_{11}+\omega_{22}}{2})+2 k\tau\ell_\mathrm{ex}^2[1-\xi(\omega_{11}+\omega_{22})]$ for the nutational linewidth and to $\Delta\omega_P(k\ll 1)\approx \frac{\alpha}{2}(\omega_{11}+\omega_{22})+2 k\tau\ell_\mathrm{ex}^2[1-\xi(\omega_{22}+\omega_{11})]$ for the precessional one. Finally, we note that by setting $\xi\rightarrow 0$ in eqs.\eqref{eq:dispersion relation precession},\eqref{eq:precession linewidth} for precessional spin waves, we recover the classical expressions $\omega_P=\sqrt{\omega_{11}\omega_{22}}-2k\tau\ell_\mathrm{ex}^2$ and $\Delta\omega_P=\frac{\alpha}{2}(\omega_{11}+\omega_{22})+2k\tau\ell_\mathrm{ex}^2$.

\begin{figure}[t]
    \centering
    \includegraphics[width=0.7\linewidth]{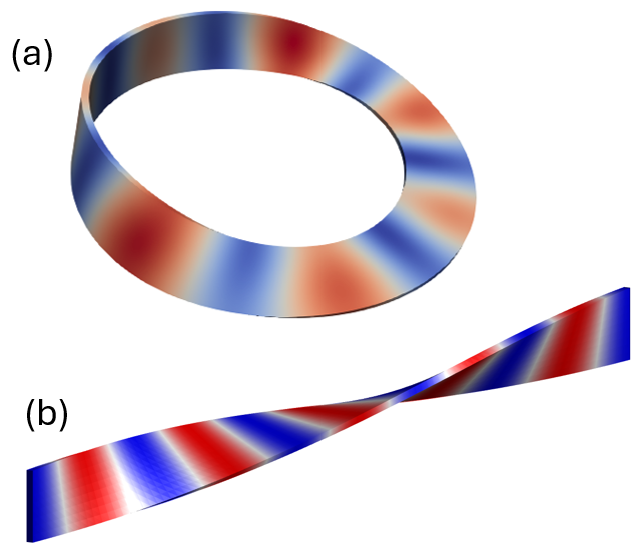}
    \caption{Examples of twisted nanostrips: (a) M\"obius; (b) Helix. The color code represents spin wave oscillation amplitude at each location ranging from zero (blue) to maximum (red).}
    \label{fig:strips}
\end{figure}
%\section{Interplay between chirality and topology}

The developed one-dimensional theory helps to understand the role of the surface twist angle on spin waves propagation, but further insights arise when one considers the case of confined nanoscale strips.

In fact, as it happens for flat strips, the boundary conditions produce quantization of the spin wave wavenumbers. Nevertheless, in the case of chiral (curved and twisted) nanostrips, boundary conditions are also affected by the topological properties of the considered surface. For instance, boundary conditions strongly depend on whether the axis of the surface is a closed or open curve. To disentangle such an interplay between chirality and topology, we consider two soft nanostrips, a M\"obius strip and a helical strip, with the same length and width $L,w$, constant curvature and torsion that, from our point of view, 
provide the two archetypes for distinct effects of twisting and topology. 

The M{\"o}bius strip can be defined 
as ruled surface when the axis is a planar circle of radius $R$ (thus having constant curvature $\kappa=1/R$) and torsion $\tau= n/(2 R)$ ($n\in\mathbb{Z}-\{0\}$ refers to positive or negative twist angle around the strip axis) as $\bm r_a(u)=R\cos(u/R)\bm e_x+R\sin(u/R)\bm e_y=R\bm n$,
where $\bm t=d\bm r_a/du=d\bm n/du$, $\bm n$ and $\bm b=\bm t\times \bm n$ are the unit-vectors tangent, normal and binormal to the axis, respectively. The local basis triple $(\bm e_1,\bm e_2,\bm e_3)$ can be computed as outlined before.
An example of M{\"o}bius strips is depicted in  Fig.\ref{fig:strips}(a).

According to eq.\eqref{eq:twist angle}, for a M{\"o}bius strip, which is a non-orientable surface for $n$ odd, the continuity conditions at the beginning and the end of the axis curve implies $\theta(2\pi R)=n\pi$,
which means that helical spin waves must fulfill the antiperiodic boundary condition $\bm\varphi_h((2\pi R,v)e^{i n\pi}=\bm\varphi_h(2\pi R,v)(-1)^n =\bm\varphi_h(0,v))$
that produces the following quantization for the wavenumber:
\begin{equation}
    2\pi k R=n\pi +2 h \pi \quad\Rightarrow \quad k_h=\frac{n+2h}{2R}\,, \, h\in\mathbb{Z} \,. \label{eq:Moebius quantization}
\end{equation}

\begin{figure}[t]
    \centering 
     \includegraphics[width=0.9\linewidth]{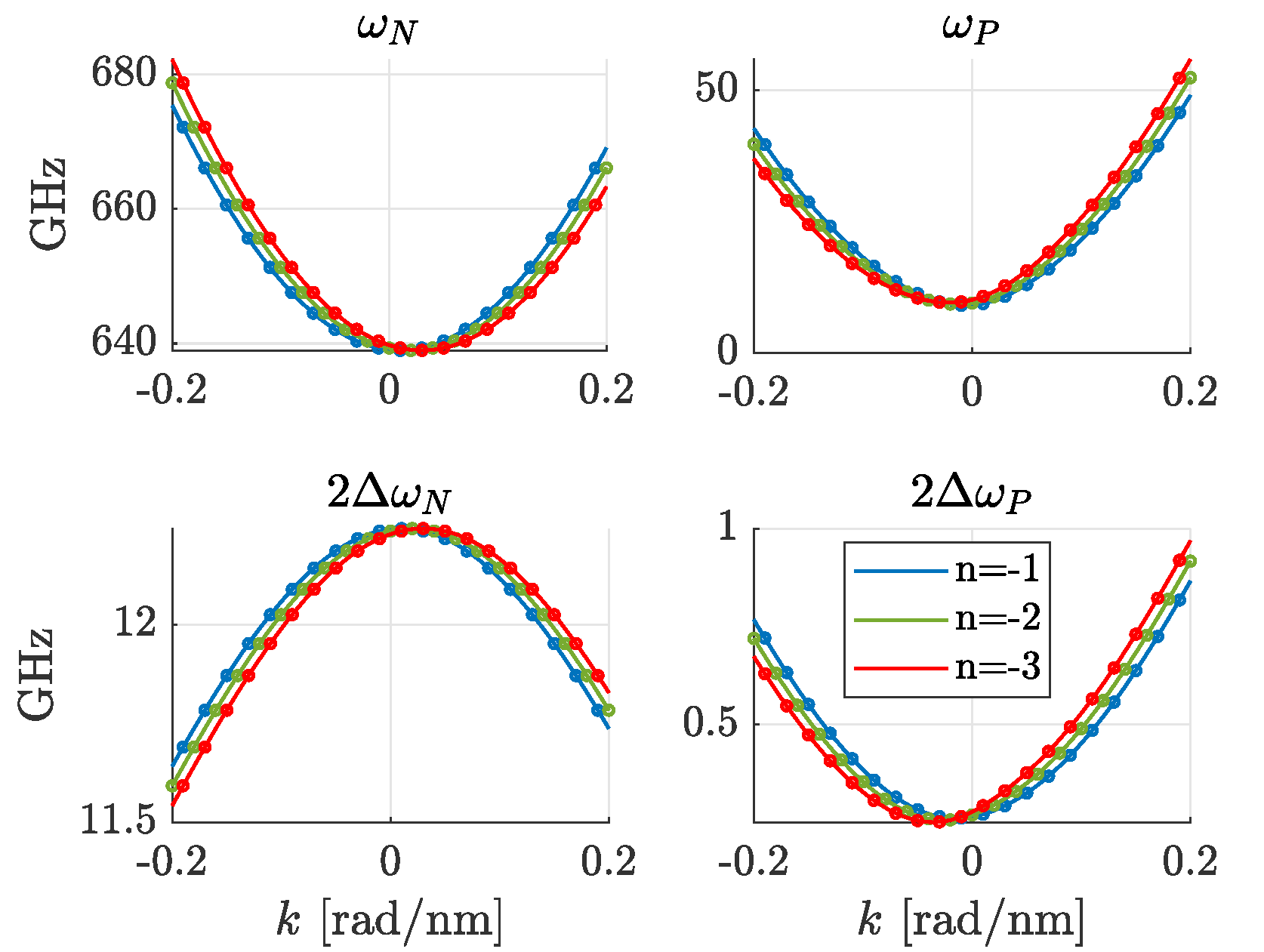}
    \caption{Dispersion relations for M{\"o}bius strips with 
   $n=-1$ (blue), $n=-2$ (green), $n=-3$ (red). 
   Solid lines rely on eq.\eqref{eq:dispersion relation nutation} (resp. eq.\eqref{eq:dispersion relation precession}) in left (resp. right) panels, symbols refer to  wavenumber quantization \eqref{eq:Moebius quantization}. 
    }
    \label{fig:comparison theory FEM Moebius strips}
\end{figure} 

We have applied the developed theory to three left-handed M{\"o}bius strips with radius $R=50$ nm, width $w=20$ nm, thickness 2 nm and increasing chirality (i.e. complete twist angle per turn) $n=-1,-2,-3$. The material parameters are those of permalloy\cite{Neeraj2021inertial}: $M_s=800$ kA/m, $A_\mathrm{ex}=13$ pJ/m, which yield $\ell_\mathrm{ex}=5.69$ nm, $t_\mathrm{in}=1.2$ ps meaning $\xi=0.045$. 

\begin{figure}[t!]
    \centering 
    \includegraphics[width=0.9\linewidth]{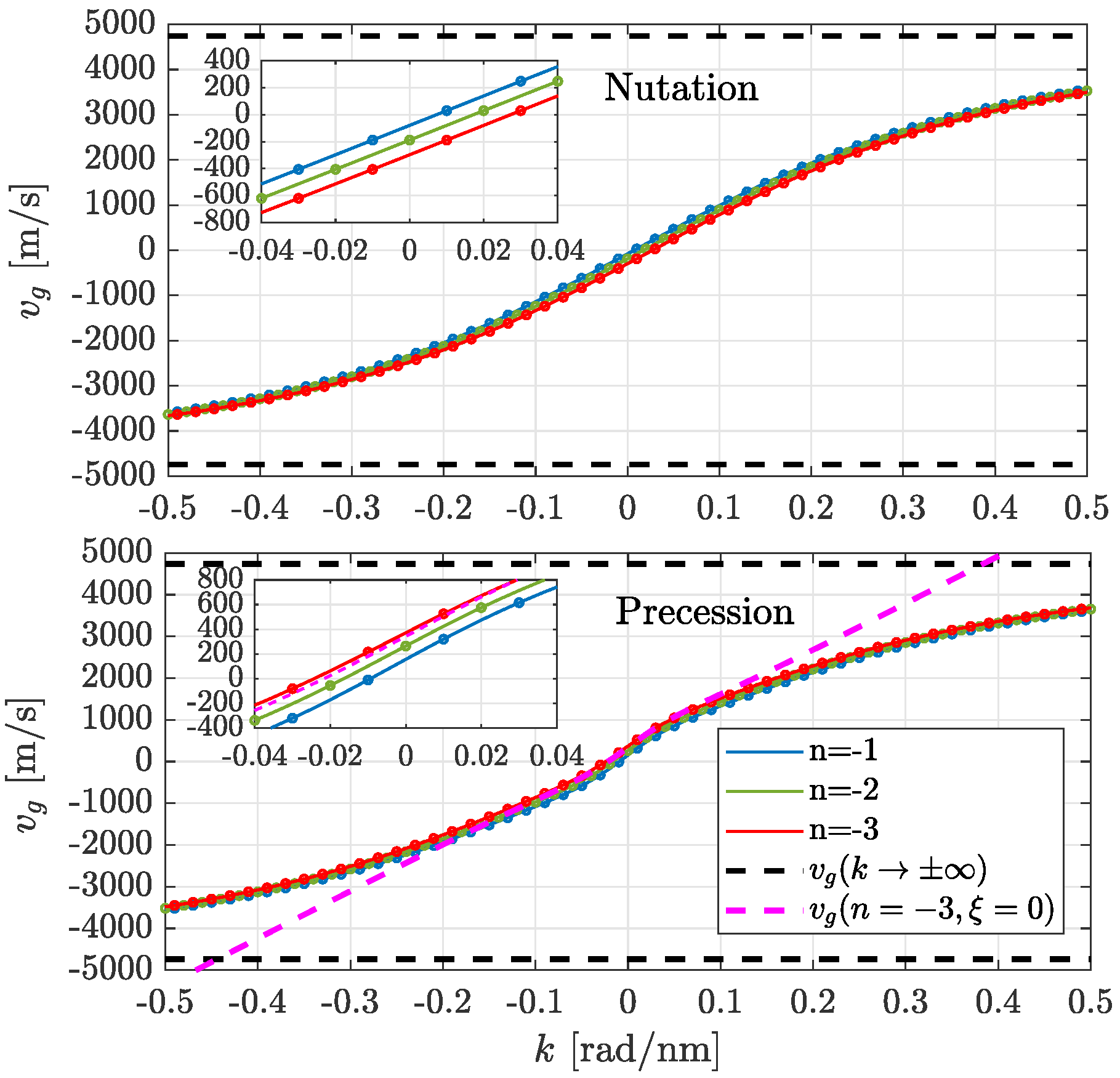} 
    \caption{Comparison between group velocities of spin waves for M{\"o}bius strips with $n=-1$ (blue), $n=-2$ (green), $n=-3$ (red). Solid lines refer to $v_g(k)=\gamma M_s\partial \omega/\partial k$, dashed black lines correspond to the ultimete speed limit $v_g(k\rightarrow\pm\infty)=\pm\frac{\ell_\mathrm{ex}}{t_\mathrm{in}}$, symbols refer to wavenumber quantization \eqref{eq:Moebius quantization}. Dashed purple line refers to 
    %eq.\eqref{eq:group velocity no inertia} 
    $v_g(k)$ with $n=-3$, $\xi=0$ (no inertia). The insets magnify the regions around $k=0$.}
    \label{fig:comparison group velocity Moebius}
\end{figure}

For the equilibrium magnetization directed along the axis of the strip, the internal field is $h_0=-\kappa^2 \ell_\mathrm{ex}^2$ while formulas 
for the demagnetizing factors\cite{aharoni1998demagnetizing} with $p=w/d=10$ 
give $N_1=0.1211, N_2=0.8789$. This means that eqs.\eqref{eq:nutation fundamental frequency}-\eqref{eq:FMR frequency} yield nutation and precession fundamental resonance frequencies
$\omega_N(k=0)\approx 22.71,22.72,22.74$ corresponding to 639.1, 639.3, 639.8 GHz and $\omega_P(k=0)\approx 0.3216,0.3357,0.3583$ corresponding to 9.050, 9.444, 10.08 GHz in physical units, respectively.   
The diagrams arising from formulas \eqref{eq:dispersion relation nutation}-\eqref{eq:precession linewidth} for dispersion relations and FWHM linewidths are reported in Fig.\ref{fig:comparison theory FEM Moebius strips} and Fig.\ref{fig:comparison group velocity Moebius} for group velocities. 
It is apparent that the non-reciprocity of spin waves arises from the nonzero torsion that produces the geometric phase accumulation.
One can clearly see that the more twisted is the strip, the more pronounced is the nonreciprocal propagation. In addition, it is evident that the group velocity exhibits substantial deviation from the classical exchange spin-wave for wavelength around 30 nm ($k\sim 0.2$, compare dashed purple and solid red line in the bottom panel of fig.\ref{fig:comparison group velocity Moebius}) and clearly attains the ultimate predicted speed limit (dashed black lines) $\pm\ell_\mathrm{ex}/t_{in}$ for large wavenumber.

The second surface is the helical strip that can be defined as ruled surface
when the axis is a straight line of length $L$ (thus having zero curvature $\kappa=0$) and torsion $\tau= n\pi/L$ (the $\pm$ refers to the positive or negative twist angle around the strip axis) as $\bm r_a(u)=u\bm e_x=u \bm t$.
An example of helical strip with left-hand chirality is depicted in Fig.\ref{fig:strips}(b)
We consider a long (ideally infinite) helical strip obtained repeating the unit-cell of length $L=2\pi R$ defined by the above equations for a large number of times (ideally infinite). Such a helix has the same torsion $\tau= n/(2R)$ or, equivalently, the same twist angle as the M{\"obius} strip analyzed in the previous section.
Nevertheless, remembering eq.\eqref{eq:twist angle},  the continuity conditions at the beginning and the end of the helix unit-cell implies $\theta(L=2\pi R)=2 n\pi$,
which means that helical spin waves must fulfill the periodic boundary condition $\bm\varphi_h(2\pi R,v)e^{i 2 n\pi}=\bm\varphi_h(2\pi R,v) =\bm\varphi_h(0,v)$
that produces the following quantization for the wavenumber:
\begin{equation}
    2\pi k R=2 n\pi +2 h \pi \quad\Rightarrow \quad k_h=\frac{n+h}{R}\,, \, h\in\mathbb{Z} \,. \label{eq:helix quantization}
\end{equation}
We stress that the quantization of wavenumbers described by eq.\eqref{eq:helix quantization} is different from that valid for M{\"o}bius strips (see eq.\eqref{eq:Moebius quantization}) even if they have the same torsion $\tau= n/(2R)$, in that they differ by a shift of $n/(2R)$ purely due to the different topology. 
Thus, the dispersion relation for helical strips will be given by eqs.\eqref{eq:dispersion relation nutation}-\eqref{eq:dispersion relation precession} setting $\kappa=0$ and using the quantization rule \eqref{eq:helix quantization}. 

In conclusion, our work establishes a comprehensive theoretical description of inertial spin-wave dynamics in twisted magnetic nanostrips, revealing how curvature, torsion, and topology interplay to produce terahertz-frequency nonreciprocal magnonic modes. The emergence of a geometric (Berry) phase, driven by inertial effects in curved geometries, provides a natural mechanism for symmetry breaking in spin-wave propagation. The derived analytical expressions for dispersion and damping capture both GHz and THz regimes, offering clear physical insights into parameter dependencies and, combined with the identification of topology-dependent quantization, open new perspectives in curvilinear THz magnonics. These findings lay the groundwork for designing chiral, high-frequency spintronic devices based on geometric control of spin-wave dynamics.

\section*{Acknowledgments}
M.d'A. acknowledges support from the Italian Ministry of University and Research, PRIN2020 funding program, grant number 2020PY8KTC. 
R.H. acknowledges support from the France 2030 government investment plan managed by the French National Research Agency ANR under grant reference PEPR SPIN – [SPINTHEORY] ANR-22-EXSP-0009.

\bibliography{paper.bib}

\end{document}